\documentclass[aps,amsmath,prc,preprint]{revtex4}
\usepackage{epsfig}
\usepackage{color}
\usepackage{slashed}
\usepackage{bbold}
\newcommand{\mn}{m_\text{\tiny N}}
\newcommand{\ecm}{E_\text{\tiny c.m.}}
\newcommand{\app}{a^C_{pp}}
\newcommand{\ann}{a_{nn}}
\newcommand{\anps}{a_{np,s}}
\newcommand{\anpt}{a_{np,t}}
\newcommand{\atni}{A_{3\text{NI}}}

\newcommand{\eftnopi}{\mbox{EFT$(\not \! \pi)$}}
\begin{document}
\title{Constraining the neutron-neutron scattering length with \eftnopi}

\author{Johannes Kirscher}
\email{kirscher@ohio.edu}
\author{Daniel R. Phillips}
\affiliation{Department of Physics and Astronomy, Ohio University, Athens, Ohio 45701, USA}

\begin{abstract} \vspace*{18pt}
We compute a model-independent correlation between the difference of neutron-neutron and proton-proton scattering lengths $|\ann-\app|$ and the splitting in binding energies between Helium-3 and tritium nuclei. We use the effective field theory without explicit pions to show that this correlation relies only on the existence of large scattering lengths in the NN system. Our leading-order calculation, taken together with experimental values for binding energies and $\app$, yields $\ann=-22.9\pm 4.1~$fm.
\end{abstract}

\maketitle

\section{Introduction}

In quantum chromodynamics (QCD) there are two effects that lead to violations of isospin symmetry. First, the electromagnetic interaction between quarks, and hence that between protons ($p$) and neutrons ($n$), does not respect the symmetry: $V^{em}_{pp}$ is different from $V^{em}_{np}$ and $V^{em}_{nn}$. Second, the difference between up and down quark masses ($m_{u,d}$) means that isospin symmetry is violated even in the absence of electromagnetic forces. The existence of \mbox{$m_d > m_u$} results in, for instance, different nucleon masses \mbox{$m_n > m_p$}, which has profound consequences for nucleosynthesis in the early universe, and nuclear physics in general~\cite{Barr}. A better understanding of
isospin-symmetry breaking is therefore of deep interest to nuclear physics, and nuclear physicists. There has been much recent progress in this direction, with experiments at TRIUMF and IUCF exploring novel signatures of the violation of isospin symmetry, see Ref.~\cite{MOS} for a review.

The effect of isospin violation is significant in the nucleon-nucleon (NN) scattering lengths.
That is because these scattering lengths are the result of fine tuning between the range and depth of the nuclear potential, and so small differences in either can lead to appreciable shifts in the scattering lengths.
In a world of an isospin-symmetry-conserving interaction, both the neutron-neutron ($nn$) and the proton-proton ($pp$) channel are degenerate with the $^1S_0$ neutron-proton ($np,s$) channel, because
for two like fermions in a relative S wave, Fermi statistics allow for a spin-singlet, total spin $J=0$ state only. Hence, even relatively weak isospin-violating interactions could have a
significant effect on the $nn/pp$ system.

The values of the proton-proton and neutron-proton scattering lengths, $\app,\anps$ are quite well established, see Table~\ref{tab.input}. For charged particles, $\app$ is observable as the
leading-order (LO) parameter in a generalized effective-range expansion that includes the effects of the non-zero Coulomb repulsion in the asymptotic states.
In contrast, $pp$ phase shifts, and hence the $pp$ scattering length, obtained exclusively from the
strong part of the $pp$ interaction retain a residual dependence on the specific model for this short-range force, \textit{i.e.},  they are dependent upon the renormalization scale and scheme~\cite{Sauer,Kong98,Gegelia03,Pavon09}.

Experiment reveals the differences between $\app$, $\anps$, and $\app$, and thus that charge independence and charge symmetry are both broken. Charge independence is associated with an arbitrary rotation in isospin space
while charge symmetry is conserved if a rotation about $\pi$ in isospin space leaves observables invariant (see, \textit{e.g.}~\cite{MOS}).
Current data allows for a relatively accurate extraction of $np$ and $pp$ scattering lengths, compared to the $nn$ parameter where the uncertainty is about two orders of magnitude larger (see Tab.~\ref{tab.input}).
For a recent, thorough review of the status of experiments see Ref.~\cite{Ga09}. Direct measurements are on the horizon~\cite{Yaguar}, but until now constraints on $\ann$ come from final-state interactions. However,
conflicting values for $\ann$ have resulted from attempts to follow this avenue in different few-nucleon reactions---indeed, conflicting values have resulted from different experiments investigating final-state
interactions in $n + d \rightarrow n + n + p$.

To extract $\ann$ accurately from this three-body deuteron-breakup experiment, the outgoing particles should predominantly be in a state of very low neutron-neutron relative momentum.
Furthermore, the proton's effect on the detected neutron pair should be minimal. This condition is satisfied for a large separation of the scattered pair from the proton which remains at rest in
the lab frame after the collision. This takes place in the $nn$ quasi-free scattering (QFS) kinematics. The cross sections corresponding to this and the analogous $np$-QFS configuration constitute the experimental input for the
subsequent extraction of $\ann$. Currently, there are two data sets from which an identical theoretical method extracts conflicting values for $\ann$. One setup records kinematical information of one
neutron and the proton~\cite{HU001,HU002} yielding $\ann=-16.1\pm0.4~$fm, while the other detects all three outgoing nucleons~\cite{GO99,GO06} and produces $\ann=-18.7\pm0.7$.
The theoretical model employs the CD-Bonn~\cite{bonn} NN potential and the charge-independent TM~\cite{tm} three-nucleon interaction.
Recently, the sensitivity of theoretical predictions for the $nn$ QFS cross section was investigated~\cite{WI10}. There it was shown that the $nn$-QFS cross section has a stronger dependence on $r_{nn}$ relative to changes in $\ann~$---after all, both experiments deal with small yet non-zero $nn$ energies, and so $r_{nn}$ would be expected to play some role. Ref.~\cite{WI10} found values of $\ann$ and $r_{nn}$ that plausibly fit both sets and would resolve the discrepancy, albeit at the expense of introducing appreciable charge-symmetry breaking in the nucleon-nucleon effective ranges. 

In view of these complications 
another way to ``measure" $\ann$ would be of great interest. Here the effective field theory without explicit pions, \eftnopi, is used to show that the difference of $nn$ and $pp$
scattering lengths, $\Delta(a):=\ann-\app$, is correlated with the trition-Helium-3 binding-energy difference, $\Delta(3):=B(t) - B(^3\text{He})$. This correlation has been known for many years within the context of models of the NN interaction. See, e.g.~Refs.~\cite{BCS78,CB87,WIS91,Pu97,MM01}, and Ref.~\cite{MI90}, which contains a review of work up until 1990. However, in these works various models, each of which produce a specific $\ann$, were used to compute $\Delta(3)$. The possibility of mapping out the general relationship and using the result to constrain $\ann$ was not explored. Here we develop the correlation between $\Delta(a)$ and $\Delta(3)$ within a leading-order \eftnopi~calculation, which shows that this correlation stems solely from the existence of large scattering lengths in the NN system.
We give an indication of how higher-order corrections can be expected to impact our result, and hence derive a constraint on $\ann$, using the value of $\app$ given in Table~\ref{tab.input}.

In general, for systems where  the scattering length $a$, is much larger than the range of the interaction $R$, an effective field theory based on the scale separation $R \ll |a|$ can be used to derive model-independent results~\cite{vK99,Ka98A,Ka98B,Ge98,Bi99,Ch99}. In nuclear physics this is \eftnopi, and it is an expansion in $R/a$, with $a$ given by the numbers in Table~\ref{tab.input}.
\eftnopi~can be used to derive ``universal" results that rely only on the existence of large scattering lengths. It has been used to compute triton ($B(t)$), Helium-4 ($B(\alpha)$), and Helium-6 binding energies~\cite{Bd99,Pl04,Ki10}.
At leading order in the $R/a$ expansion there are three parameters in the EFT that then yield predictions for all other observables in systems with $A \leq 4$: these can be taken to be the spin-singlet and spin-triplet NN scattering lengths, and the binding energy of the three-nucleon system. At next-to-leading order (NLO)  in the $R/a$ expansion the NN effective ranges enter the problem, with all other
low-energy NN parameters only affecting answers beyond NLO~\cite{HM01,Bd02,Pl06,Ki09}.
This specifies the general EFT prescription of fitting a minimal set of LECs to observables in order to make predictions for all other observables correlated to the input set.

In particular, \eftnopi~provides a map from an input set, \textit{e.g.}, the scattering length $\anpt$, to a correlated set, whose elements, \textit{e.g.},
the deuteron binding energy $B(d)$, are predicted with known theoretical uncertainty. However, even if this set of correlations is well mapped out in the $A$-body system there appears to be no rigorous way to determine {\it a priori} whether $A+1$-body observables will also be correlated with the $A$-body input quantities. 
For example, the triton binding energy $B(t)$ is {\it not} correlated with the $np$ singlet and triplet scattering lengths $\{\anps,\anpt\}$~\cite{Bd99},
while the binding energy of the $\alpha$ particle {\it is} correlated
with $\{\anps,\anpt,B(t)\}$~\cite{Pl04}. For $B(t)$, a strong sensitivity to short-distance structure---parameterized, for instance, by a momentum-space cutoff $\Lambda$---is found, while the dependence of $B(\alpha)$ on $\Lambda$ is
parametrically small once the value of $B(t)$ is fixed. This latter phenomenon, known as the Tjon line~\cite{tjon}, allows for a prediction of the $\alpha$-particle binding energy once that of the triton is known.

However, once it has been established that a higher-$A$ observable is a member of the set of correlated quantities, we may use that observable to constrain properties of smaller subsystems.
This prescription is used here, where we consider the binding energy difference $\Delta(3)$, which would be zero if isospin were an exact symmetry of nature. We therefore exploit another feature of the EFT, namely the absence of a qualitative hierarchy amongst low-energy observables in their role as input to fix the LECs.
Thus, our interaction will take $\{\anps,B(d),\app,B(t),B(^3\text{He})\}$ ($A\leq3$) as input, and we will obtain the scattering length $\ann$ ($A=2$) as output.
Analogously, the Tjon line could be used to predict a range of $B(t)$ values which are consistent with $B(\alpha)$.

Our calculation of $\Delta(3)$ considers isospin violation from Coulomb interactions, and from the difference in NN scattering lengths.Experimentally, $\Delta(3)$ is known to be $764~$keV~\footnote{Numerical values for physical binding energies are taken as referenced in the TUNL database,
\texttt{http://www.tunl.duke.edu/NuclData/}, access date: April $2011$}.
Note that we do not claim that our leading-order \eftnopi~calculation of the individual tri-nucleon binding energies is 
this accurate,
but we are examining a binding-energy difference that would be zero in the symmetry limit, and so a leading-order calculation of the difference already provides a useful constraint on $|\ann-\app|$.
In pursuing such a calculation we are, though, implicitly assuming that any isospin-violating component of the three-nucleon force in \eftnopi~enters only at sub-leading orders.
We will present evidence that supports this assumption.

The history of analyses of the impact of charge-independence breaking (CIB) and charge-symmetry breaking (CSB) on $\ann-\app$ and the trinucleon binding-energy splitting is a rich one (see \textit{e.g.} \cite{MI90} for a review
of most of the investigations predating the advent of EFT methods in few-nucleon theory). Today, modern high-precision nucleon-nucleon force models predict a trinucleon binding-energy difference,
$\Delta(3)$, in good agreement with experiment: $\Delta(3,\texttt{AV18+UIX})=756(1)~$keV~\cite{Pu97} and $\Delta(3,\texttt{AV18})=762(9)~$keV~\cite{MM01}. Even though the individual binding energies of the triton and Helium-3 receive significant contributions from a three-nucleon interaction (TNI) in these models, $\Delta(3)$ is driven by the difference in
the nucleon-nucleon scattering lengths---at least once electromagnetic effects are properly accounted for. Isospin-violating TNIs were considered in the
framework of chiral perturbation theory in Refs.~\cite{KO05,EMP05}, and the leading-order isospin-violating TNI was found to contribute approximately $5~$keV to $\Delta(3)$~\cite{KO05}. 

A recent LO analysis of $\Delta(3)$ in \eftnopi~included the Coulomb interaction nonperturbatively. With $\ann$ and $\app$ as input, a value of $\Delta(3)=0.82~$MeV was predicted using Faddeev methods and the dibaryon formalism~\cite{AB10}.
In Ref.~\cite{HK11}, the authors applied \eftnopi~at LO, NLO, and N$^2$LO to predict proton-deuteron scattering- and bound-state observables. Their comprehensive analysis demonstrates the
usefulness of the EFT prescription in low-energy $pd$ scattering and in the Helium-3 bound state, since an order-by-order decrease of the theoretical uncertainty is obtained. In Ref.~\cite{HK11} too,
$\ann$ and $\app$ were used to fit the low-energy constants of the EFT. Neither of these \eftnopi~analyses included isospin-violating TNI.
\par
The article continues with an introduction of \eftnopi~as the theory underlying the interaction potential. Next, a section on the numerical method of a refined version of the resonating group method (RGM),
used to solve the few-body problem, precedes the presentation of the results. The results section includes subsections discussing the uncertainty estimates due to suppressed higher-order long-
and short-range interactions, and the limiting (hypothetical) case of $\ann\to\app$. We then offer our conclusions, before assessing the numerical stability of the refined RGM in an appendix.

\section{Input: \eftnopi~with leading-order Coulomb interactions and isospin violation}

The effective field theory without explicit pions (\eftnopi) at leading order in the expansion parameter $Q/M$ is defined by the Langrangean
\begin{eqnarray}\label{eq.lo-lagrangean-sy}
\mathcal{L}^{(\text{CI})}&=&N^\dagger\left(i\partial_0+\frac{\vec{\nabla}^2}{2\mn}\right)N
+C_1(N^\dagger \sigma_2\tau_2\tau_i N)\cdot(N^\dagger \sigma_2\tau_2\tau_i N)\nonumber\\
&&+C_2(N^\dagger \sigma_2\sigma_i\tau_2 N)\cdot(N^\dagger \sigma_2\sigma_i\tau_2 N)
+C_{3\text{NI}}(N^\dagger N)(N^\dagger\tau_i N)(N^\dagger\tau_i N)\;\;\;,
\end{eqnarray}
where the six-nucleon contact term renormalizes the $S=1/2$ nucleon-deuteron channel (see, \textit{e.g.} \cite{Bd99}).
The (iso)spin matrices ($\vec{\tau}$)$\vec{\sigma}$, with indices specifying the Cartesian component, project onto spin singlet and
triplet with the respective low-energy constant, $C_{1,2}$.
The two-nucleon amplitude derived from this Lagrange density matches the effective range expansion.
The description is appropriate if the typical momentum exchange $Q$ between interacting nucleons of mass $\mn$ is small relative to the high-energy scale $M\approx m_\pi$.
For $Q\gtrsim m_\pi$, this theory is not applicable as is uses the neutron and proton
Pauli spinors \mbox{$N=\left(\vert p,s=1/2 \rangle\,,\,\vert n,s=1/2 \rangle\right)$} as degrees of freedom---nothing else.
The interaction in Eq.~(\ref{eq.lo-lagrangean-sy}) is charge independent and does not discriminate between neutron-neutron ($nn$), proton-proton ($pp$), and proton-neutron ($pn$)
pairs in the $^1S_0$ NN channel. For a comprehensive analysis of systems including at least two charged protons at low energies, the effect of
the electromagnetic force cannot be neglected. This is apparent in the measured difference between the $pp$ and $np$ $^1S_0$ scattering lengths,
$\app=-7.8063\pm0.0026~$fm~\cite{pp-pwa} compared to $\anps=-23.748\pm0.009~$fm~\cite{np-a}, where Eq.~(\ref{eq.lo-lagrangean-sy}) yields
$\app=\anps$.
\par
Electromagnetic interactions are considered canonically by promoting the Lagrangean Eq.~(\ref{eq.lo-lagrangean-sy}) to a local
gauge theory, invariant under local U(1) transformations. Using Coulomb gauge $\vec{\nabla}\cdot\vec{A}=0$, the contributions of the
gauge fields $A_\mu$ to the $pp$ amplitude can be split into a part which scales as $\alpha/Q^2$, resulting from the part of
the covariant derivative proportional to $N^\dagger eA_0N$ (``Coulomb photons''), and others which are either suppressed by powers
of $Q/M$ or at least $\mn^{-2}$ (``transverse photons''). Here, this estimate justifies the usage of the Coulomb potential
resulting from the ``exchange'' of one Coulomb photon to account for the electromagnetic interaction at low energies.
\par
Without charge-independence-breaking (CIB) mechanisms stemming from a broken flavor SU(2) symmetry in the u-d quark sector, \textit{i.e.},
$m_u\neq m_d$, the $np$ and $nn$ $^1S_0$ channels would still be degenerate. To refine the
analysis, the lowest-order contribution from this asymmetry is included. The dominating terms are expected to be of lowest-mass dimension
while a dependence on the direction of the isovector is now admissible in order to distinguish $nn$, $pp$, and $np$ vertices. In a hierarchy of
isospin-violating interactions~\cite{VK96,Wa01}, those contact terms which are expected to scale as $\epsilon Q^0$ with
$\epsilon=\frac{m_d-m_u}{m_d+m_u}\approx\frac{1}{3}$ should be subleading compared to the Coulomb potential. We include both the Coulomb potential, and these isospin-violating short-range operators, in our calculation, allowing us to formulate a model-independent assessment of which $\ann$ values are
consistent with the experimental tri-nucleon binding energy splitting $\Delta(3)$,
the singlet(triplet) $np$ S-wave scattering lengths $a_{\text\tiny np,s(t)}$, and $\app$. 

The subsequent analysis is therefore based on
the Lagrangean (\ref{eq.lo-lagrangean-sy}), including pieces obtained by the minimal substitution $\partial_0 \rightarrow \partial_0 + i e A_0$, combined with explicit CSB terms:
\begin{eqnarray}\label{eq.lo-lagrangean-asy}
\mathcal{L}^{(\text{CSB})}&=& C^{nn}_S \left(N^\dagger N\right) \left(N^\dagger \sigma_2\left(\mathbb{1}+i\tau_2\tau_1\right)N\right)
+ C^{pp}_S \left(N^\dagger N\right) \left(N^\dagger \sigma_2\left(\mathbb{1}-i\tau_2\tau_1\right)N\right).
\end{eqnarray}
The tree-level diagrams corresponding
to these interactions define the isospin-violating part of the potential
\begin{eqnarray}\label{eq.pot-vertex}
\hat{V}^{(\text{CSB})}&=&\sum_{i<j}^A\Big[\left(\frac{e^2}{4\vert\vec{r}(i)-\vec{r}(j)\vert}+C^{pp}_Sf_\Lambda(\vec{r}_{ij})\right)\left(1+\tau_3(i)\right)\left(1+\tau_3(j)\right)\nonumber\\
&&+C^{nn}_S\left(1-\tau_3(i)\right)\left(1-\tau_3(j)\right)f_\Lambda(\vec{r}_{ij})\Big]\frac{1}{4}\left(1-\vec{\sigma}(i)\cdot\vec{\sigma}(j)\right)
\end{eqnarray}
in coordinate representation with $f_\Lambda(\vec{r}_{ij}):=\left(\frac{\Lambda^3}{8\pi^{3/2}}\right)e^{-\frac{\Lambda^2}{4}\vec{r}_{ij}^2}$. Meanwhile, the isospin-conserving piece of the potential, which stems from Eq.~(\ref{eq.lo-lagrangean-sy}), is:
\begin{eqnarray}\label{eq.pot-np}
\hat{V}^{(\text{CI})}&=&\sum_{i<j}^A f_\Lambda(\vec{r}_{ij}) \left[\frac{1}{2}\left(1-\vec{\sigma}(i)\cdot\vec{\sigma}(j)\right) C_1 + \frac{1}{2}\left(3+\vec{\sigma}(i)\cdot\vec{\sigma}(j)\right) C_2\right]\nonumber\\
&&+\sum\limits_{\stackrel{i<j<k}{\text{cyclic}}}^Af_\Lambda(\vec{r}_{ij})\cdot f_\Lambda(\vec{r}_{jk})C_{3\text{NI}}\,\vec{\tau}_i\cdot\vec{\tau}_j\,\,.
\end{eqnarray}
These operators $\hat{V}^{(\text{CSB})}$, $\hat{V}^{(\text{CI})}$, are used in the Schr\"odinger equation and originate from the momentum-independent vertices via a Gaussian regulator of the non-separable form
$f_\Lambda(\vec{p},\vec{p'})=e^{-\left(\vec{p}-\vec{p'}\right)^2/\Lambda^2}$, which Fourier transforms into a Gaussian depending
on the relative coordinate.
From this non-relativistic equation of motion, variational
approximations to the bound and scattering states of the two- and three-nucleon systems are obtained. In solving the equation of motion,
the potential is iterated.

The iteration of the unregulated ($\Lambda\to\infty$) interaction $\sim C_2$ in the ${}^3$S$_1$ channel yields a $np$ total isospin $T=0$ amplitude~\cite{We91}:
\begin{equation}
T^{T=0,np}(p) =\left[\frac{1}{C_2} - \int \frac{d^3q}{(2 \pi)^3} \frac{1}{E + i \epsilon - q^2/\mn}\right]^{-1}
\end{equation}
where $p=\sqrt{\mn E}$ is the relative momentum of the $np$ pair in the c.m. frame and $\epsilon$ is a positive infinitesimal. $C_2$ can then be chosen in order to obtain the real bound state (deuteron) in that channel:
\begin{equation}
T^{T=0,np}(p) =\frac{4 \pi}{\mn}\frac{1}{\gamma + i p},
\label{eq.Tdeut}
\end{equation}
where we have chosen to ignore terms which are suppressed by $p^2/\Lambda$ in the denominator, although these pieces will be present in the denominator for any finite $\Lambda$, and will thus produce a non-zero, positive~\cite{PC97}, and cutoff-dependent effective range. The binding momentum is denoted here as $\gamma=\sqrt{\mn B(d)}$.

Similarly, the iteration of either $C_1$ or $C_1 + C_S^{nn}$ produces the virtual bound states
in the $nn$ and $np$ systems:
\begin{equation}
T^{T=1,np/nn}(p)=\frac{4 \pi}{\mn}\frac{1}{\frac{1}{a_{np/nn}}+ i p},
\end{equation}
under the same conditions as in Eq.~(\ref{eq.Tdeut}).
The isospin-violating terms account for the difference in the scattering length of $nn$,
relative to $np$. Without the $C^{nn}_S$ counterterm, the spin-singlet $nn$ and $np$ channel would be degenerate:
$\ann=\anps$.

\begin{figure}
\includegraphics[width=0.9\columnwidth]{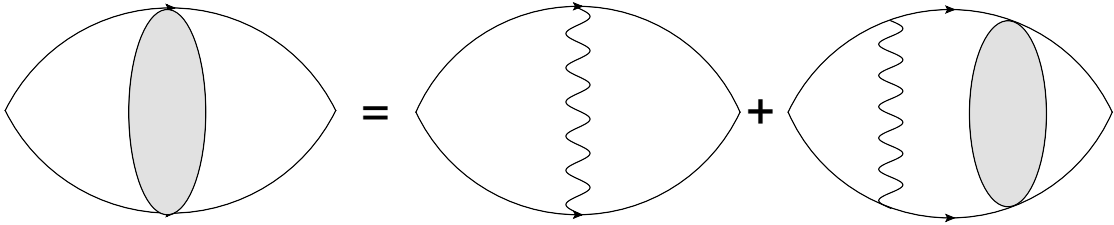}
\caption{\label{fig.coul-bubble}\small A diagrammatic definition of the Coulomb bubble $J_0$. The solid lines are nucleons, and the wavy line represents a single insertion of the Coulomb potential. The shaded blob is the t-matrix found by iterating the Coulomb potential alone, $T_{Coul}$. }
\end{figure}

The case in the proton channel is different, due to the presence of the Coulomb interaction there. The computation of $pp$ scattering was carried out to LO in \eftnopi~for S-wave NN scattering in Ref.~\cite{Kong98,KR99}. Our presentation of $pp$ scattering rests on that treatment. 
The final result is \cite{GoldbergerWatson}
\begin{equation}
T=T_{NC} + T_{Coul},
\label{eq.sum}
\end{equation}
with $T_{Coul}$ the amplitude for scattering due to the Coulomb potential alone. $T_{NC}$ encodes the purely strong scattering {\it and} the Coulomb-nuclear interference:
\begin{equation}
T_{NC}=C_\eta^2 \exp(2 i \sigma_0(\eta))\frac{1}{\frac{1}{C_1 + C^{pp}_S} - J_0(p)}.
\label{eq.TNC}
\end{equation}
Here $C_\eta^2$ is the Sommerfeld factor:
\begin{equation}
C_\eta^2=\frac{2 \pi \eta}{e^{2 \pi \eta}-1},
\end{equation}
and the Coulomb parameter $\eta :=\frac{\mn \alpha}{2 p}$, with $\sigma_l(\eta)={\rm arg} \Gamma(l + 1 + i \eta)$ ($\Gamma$ is the Euler gamma function). In Eq.~(\ref{eq.TNC}) 
$J_0(p)$ is the Coulomb-modified bubble, depicted in Fig.~\ref{fig.coul-bubble}. If computed in PDS~\cite{Ka98A} at a renormalization scale $\mu$, its finite part is:
\begin{equation}
J_0^{\rm finite}(p)=-\frac{\alpha \mn^2}{4 \pi}\left[H(\eta) - \ln\left(\frac{\mu \sqrt{\pi}}{\alpha \mn}\right) - 1 + \frac{3}{2} C_E\right] - \frac{\mu \mn}{4 \pi},
\label{eq.J0finite}
\end{equation}
once divergences in $D=4$ and $D=3$ have been dropped. In Eq.~(\ref{eq.J0finite}), the function
\begin{equation}
H(\eta)=\psi(i\eta) + \frac{1}{2 i \eta} - \ln(i\eta),
\end{equation}
with $\psi$ the derivative of the Euler Gamma function, Euler's constant $C_E=0.5772 (...) $, and $\alpha=e^2/4 \pi$ (see also~\cite{BB02,AB08,Hi08}). 

%----------------------------------------------------------------------------------------------------------------------
%----------------------------------------------------------------------------------------------------------------------

The denominator in Eq.~(\ref{eq.TNC}) can be matched to the modified effective-range expansion:
\begin{equation}
T_{NC}=-C_\eta^2 \frac{4 \pi}{\mn} \frac{\exp(2 i \sigma_0)}{-\frac{1}{\app} + \frac{1}{2} r_0 p^2 -\alpha \mn H(\eta)}.
\end{equation}
This expression for $T$ encodes the fact that the spherical Bessel functions, the asymptotic solutions for the $np/nn$ system, are now
replaced by Coulomb functions~\cite{GoldbergerWatson}. When this matching is performed at leading order ($r_0=0$) we find that 
\begin{equation}
\frac{1}{C_1+C_S^{pp}}=\frac{\mn}{4 \pi \app} + \frac{\alpha \mn^2}{4 \pi}\left[\ln\left(\frac{\mu \sqrt{\pi}}{\alpha \mn}\right) + 1 - \frac{3}{2} C_E\right] - \frac{\mu \mn}{4 \pi}.
\end{equation}
To produce the experimentally observed $\app$ the combination $C_1+C^{pp}_S$ is fitted to the value listed in Table~\ref{tab.input}. Strictly speaking, the $\app$ listed there
is extracted from data using an effective-range expansion that considers a refined version of the Coulomb potential and vacuum-polarization effects~\cite{pp-pwa}. However, here we have a long-range part of the $pp$ potential consisting only of the Coulomb interaction. The uncertainty due to higher-order long-range interaction
effects, such as vacuum polarization, is included in the overall theoretical error estimate.

Regardless of this detail though, we see that $\app$ can be obtained from the $pp$ amplitude in a manner that is independent of the renormalization scale $\mu$.
In contrast, attempting to ``switch off" Coulomb interactions and compute the effect obtained solely from the strong potential, $C_1 + C^{pp}_S$, yields a result that, within PDS, is dependent upon the renormalization scale $\mu$. $C_1 + C^{pp}_S$ contains a $\ln \mu$ piece, and there is no corresponding $\ln \mu$ piece of the Coulomb-less loop function to cancel that. Our calculation is not done using PDS, but instead with a Gaussian regulator, yielding approximately a $\ln \Lambda$ dependence of the $C^{pp}_S$ that reproduces $\app$~(the $\ln\Lambda$
dependence is seen explicitly with a sharp momentum cutoff~\cite{Kong98}).
But, in any case, this cutoff/renormalization-scale dependence renders quoting a ``strong $pp$ scattering length" a questionable exercise~\cite{Gegelia03,Pavon09}---at least
in the absence of an agreed upon choice for the regularization and renormalization scheme and scale.
Therefore, in what follows we quote all results in terms of $\app$, which is a physical observable, and as such is independent of these choices.

In practice, the equation
\begin{equation}\label{eq.sgl}
\left(\sum_{i=1}^A\frac{\vec{\nabla}_i^2}{2\mn}+\hat{V}_\text{\tiny Coulomb}+\hat{V}\left(C_{1,2},C^{nn,pp}_S,C_{3\text{NI}}\right)\right)\vert\psi\rangle =E\vert\psi\rangle
\end{equation}
is solved for two-body scattering ($E>0\;,A=2$) and three-body bound states ($E<0\;,A=3$, see ch.~\ref{sec_rgm} below).
In total, five LECs are fitted to low-energy data according to Table~\ref{tab.input}. Again, it is the sum $C_1+C^{nn}_S$ that controls the $nn$ channel, not just $C^{nn}_S$!

Isospin-violating interactions at the quark level imply furthermore a mass difference between the neutron and the proton.
This consequence is a higher-order effect in our counting, as it is not enhanced by the fine-tuning in the NN system. Hence, it is not accounted for in the calculation below, where throughout $m_n=m_p=:\mn=\frac{1}{2}(938.211+939.505)~\text{MeV}$ is used.
In sec.~\ref{subsec.un2}, the contribution of this mass difference to $\Delta(3)$ is included in the error assessment.
\section{The resonating-group method}\label{sec_rgm}
For the solution of the two- and three-body problem the variational Resonating Group Method (RGM) is employed.
In the following, the RGM is introduced
by the specific example of the variational space used for the triton calculation of this work.
\par
The ground state is expanded in two different variational bases which differ in the angular-momentum coupling scheme between the
spin- and coordinate-space components of the wave function.
First, the ansatz for the three-body state in the ``LSJ-scheme'' (total angular momentum $J$ and parity $\pi$), used here
in Jacobi-coordinate space ($\vec{\rho}_1,\vec{\rho}_2$), reads
\begin{equation}\label{eq.rgm-lsj-ansatz}
\big\vert\psi(\vec{\rho}_1,\vec{\rho}_2),J=1/2,\pi=+\big\rangle_{LS}=
f_1(\vec{\rho}_1,\vec{\rho}_2)\vert d-n\rangle+f_2(\vec{\rho}_1,\vec{\rho}_2)\vert \mbox{d\hspace{-.55em}$^-$}-n\rangle+f_3(\vec{\rho}_1,\vec{\rho}_2)\vert(nn)-p\rangle\;\;\;,
\end{equation}
namely, a linear combination of all possible two-fragment substructures within the triton: deuteron (d),
spin-singlet deuteron (\mbox{d\hspace{-.55em}$^-$}), and dineutron ($nn$). Thus, the coupling schemes of the spin (S) and isospin (T)
angular momenta are fixed, \textit{e.g.},
\begin{eqnarray}\label{eq.dn-in-lsj}
&&\vert d-n\rangle=\left[\Big[\vert\text{n},m_{s_1}\rangle\otimes\vert\text{p},m_{s_2}\rangle\Big]^1\otimes\vert\text{n},m_{s_3}\rangle\right]^{\frac{1}{2}m_S}\nonumber\\
&&
\qquad \cdot
\left[\Big[\vert\text{n},m_{t_1}=-{\frac{1}{2}} \rangle\otimes\vert\text{p},m_{t_2}={\frac{1}{2}} \rangle\Big]^0\otimes\vert\text{n},m_{t_3}=-{\frac{1}{2}} \rangle\right]^{\frac{1}{2}m_T}.
\end{eqnarray}
The square brackets are shorthand for a basis where the irreducible representations are labeled by quantum numbers corresponding
to $(S+S')^2$ for two general tensor operators with spherical components $m(m')$,
$\left[S^{m}\otimes S'^{m'}\right]^{JM}=\sum_{m,m'}(SmS'm'|JM)S^{m}S'^{m'}$ (see \textit{e.g.}~\cite{edmonds}).
The coordinate-space wave functions $f_i$ are expanded in a Gaussian basis of dimension $D_i$:
\begin{equation}\label{eq.gauss-exp-coord-wfkt}
f_i=\sum_{j=1}^{D_i}c_{ij}\exp(-\gamma_1^{ij}{\vec{\rho}_1}^{\, \, 2}-\gamma_2^{ij}{\vec{\rho}_2}^{\, \, 2})\cdot
\left[\mathcal{Y}_0\left(\vec{\rho_1}\right)\otimes \mathcal{Y}_0\left(\vec{\rho_2}\right)\right]^{L=0}.
\end{equation}
Here the two solid harmonics~\cite{edmonds} are both chosen to correspond to relative angular momentum zero. As in the Faddeev approach of~\cite{Bd99,HK11} with LO \eftnopi,
the dynamics in the variable $\vec{\rho}_1$ are affected by S-wave NN interactions only. The $\vec{\rho}_1$ coordinate parameterizes the two-nucleon fragment, for instance, the
relative coordinate between a proton and a neutron in a deuteron-neutron configuration.
As the lowest-lying bound state will occur for an angular momentum on the second coordinate, \textit{e.g.}, between the center of mass of a deuteron cluster and the remaining neutron,
which is zero as well, the total orbital angular momentum is also zero ($L=0$).
The impact of higher angular momenta on the solution for the 3N bound state is assessed in Appendix~\ref{sec.rrgm.numstab}.

The relative size of the variational parameters $c_{ij}$ then determines the overlap of the corresponding configuration with the triton,
and hence the significance of a certain combination of width parameters $\gamma_{1,2}^{ij}$ for the grouping $i$. The triton ground state, for example,
is found to have the largest overlap with the d-n grouping. To this configuration, parameters $\gamma_1$ that resemble the spatial
extent of the deuteron and $\gamma_2$ that place the neutron at a farther distance, contribute most. For the other two groupings,
$i=2,3$, the two-body fragment is unbound. In consequence, basis vectors with smaller widths $\gamma_1$, \textit{i.e.}, broader spatial extent,
become important.
\par
In the basis (\ref{eq.rgm-lsj-ansatz}), the individual spin- and orbital angular momenta are coupled separately to their total $S$ and $L$
before they combine to total $J$. This coupling scheme is different from the one used in~\cite{HK11}. The RGM \textit{ansatz}
used for two-fragment scattering resembles that calculation:
\begin{equation}\label{eq.rgm-ch-ansatz}
\big\vert\psi(\vec{\rho}_1,\vec{\rho}_2),J=1/2,\pi=+\big\rangle_{\text{ch}}=
\left[f_{1,l_\text{\tiny rel}}\otimes\Big[\vert d\rangle\otimes\vert n\rangle\Big]^{S_\text{ch}}\right]^J+
\left[f_{2,l_\text{\tiny rel}}\otimes\Big[\vert \mbox{d\hspace{-.55em}$^-$}\rangle\otimes\vert n\rangle\Big]^{S_\text{ch}}\right]^J\;\;\;,
\end{equation}
with $f_{i,l_\text{\tiny rel}}=\sum_{j=1}^{D_r}c_{ij}e^{-\gamma_{ij}^2\vec{\rho}_2^2}\mathcal{Y}_{l_\text{\tiny rel}}\left(\vec{\rho_2}\right)$
and a two-fragment wave function determined within the same framework. Now, the deuteron's $J=1$ is combined with the neutron to a channel spin
$S_\text{ch}$ which couples to the relative orbital angular momentum $l_\text{\tiny rel}\stackrel{\text{\scriptsize here}}{=}0$ to the total $J$.
With the two-fragment wave function fixed, 
the variational parameters $c_{ij}$ determine the relative
importance of the components and the expansion of the coordinate-space wave function of the third particle relative to the center of mass
of the pair.
This resembles the usage of the various dibaryon propagators $\Delta_{d(t)}^{ij(AB)}$ for the
deuteron ($^1S_0$) channel (see~\cite{Bd99,HK11}) in the Faddeev approach, and the corresponding projection to deuteron-neutron (for example) relative angular momentum zero.
The expansion parameters $c_{ij}$ of the triton in both the LS and the channel basis are obtained by solving the generalized
Eigenvalueproblem for a Hamiltonian incorporating the potential of Eq.~(\ref{eq.pot-vertex}), {\it i.e.} Eq.~(\ref{eq.sgl}).
%--
\begin{table*}
\renewcommand{\arraystretch}{1}
  \caption{\label{tab.input}{\small Data input for the LECs which specify the interaction Eq.~(\ref{eq.pot-vertex}). The values were fitted
nonperturbatively with the RGM and include the Coulomb potential's contribution. Four sets of LECs corresponding to
$\Lambda=400,800,1600~$MeV were adopted. The two sets for $\Lambda=400~$MeV correspond to different input data.}}
\footnotesize
\begin{tabular}{c|c|l}
\hline
channel & low-energy constants (LEC) & \hspace{.2cm}constraining observable\\
\hline\hline
$(np)$ $^1S_0$ & $C_1$ &\hspace{.2cm}$\anps=-23.748\pm0.009~$fm~\cite{np-a}\\
$(np)$ $^3S_1$ & $C_2$ &\hspace{.2cm}$\anpt=5.4194\pm0.002~\text{fm}$\\
             &       &\hspace{.2cm}$B(d)=2.224575\pm0.000009~$MeV~\cite{d-prop}\\
$(pp)$ $^1S_0$ & $C_1+C^{pp}_S$ &\hspace{.2cm}$\app=-7.8063\pm0.0026~$fm~\cite{pp-pwa}\\
$(npp)$ $^2S_1$ & $C_1+C^{pp}_S,C_1,C_2,C_{3\text{NI}}$ &\hspace{.2cm}$B(^3\text{He})=7.718109\pm0.000010~$MeV\\
$(nnp)$ $^2S_1$ & $C_1+C^{nn}_S,C_1,C_2,C_{3\text{NI}}$ &\hspace{.2cm}$B(t)=8.481855\pm0.000013~$MeV\\
\hline
    \end{tabular}
\end{table*}
%--

\section{Results}

Here, to obtain a result for $\ann$ different from the values of the other two-nucleon scattering lengths, the triton binding energy $B(t)$ is used as input for $C^{nn}_S$. The triton constitutes the next larger system where an interaction between two neutrons is observable. If there are no additional three-nucleon operators which affect low-energy modes and break isospin symmetry then $B(t)$ and $\ann$ are correlated, and therefore $\Delta(3)$ is correlated with $\Delta(a)$. An interaction given by Eq.~(\ref{eq.pot-vertex})
with $B(t)$ as input for $C^{nn}$ will then predict $\ann$ within our uncertainty bounds. 

\begin{figure}
\includegraphics[width=0.9\columnwidth]{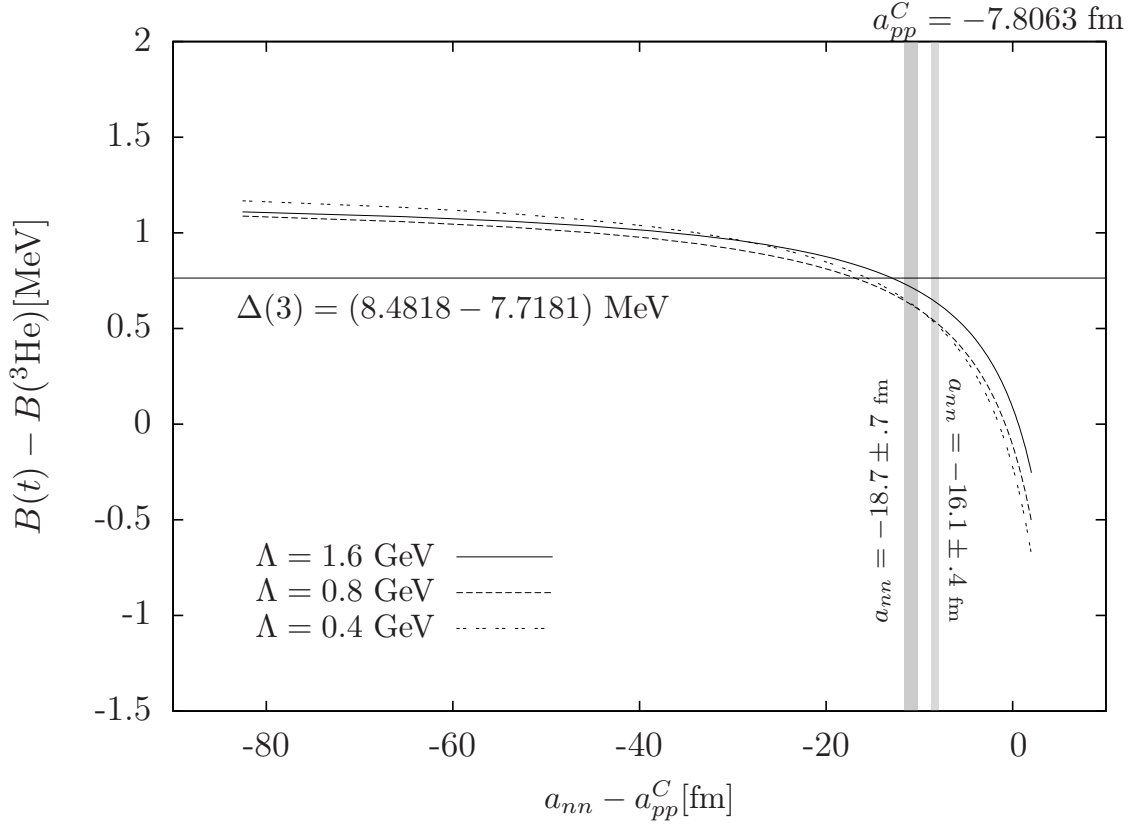}
\caption{\label{fig.res.db3}\small Correlation between the trinucleon-binding-energy difference $B(t)-B(^3\text{He})$ and the $nn$ and $pp$
scattering lengths at LO in \eftnopi. Correlation lines correspond to regulator values $400~$MeV, $800~$MeV, and $1.6~$GeV. For
$\Lambda=700~$MeV, the correlations overlap with the displayed lines and are not shown.}
\end{figure}

The results presented in Fig.~\ref{fig.res.db3} are RGM solutions to the Schr\"odinger equation with an interaction given
in Eq.~(\ref{eq.pot-vertex}) and LECs constrained by the conditions defined in Table~\ref{tab.input}. The correlation as shown
there is obtained by a variation of $C^{nn}$. The predictions for $\ann$ of four interactions, each containing different short-distance physics (see below),
inferred from the intersection of the computed correlation line with the experimental value for $\Delta(3) :=B(t)-B({}^3{\rm He})$, suggest
\begin{equation}\label{eq.res.ann}
\lim_{\Lambda\to\infty}\ann(\Lambda)\approx -20.8~\text{fm}.
%mm-nb interpolate-ann.nb
\end{equation}

An assessment of whether the chosen set of input quantities and leading-order calculation are such that this prediction discriminates between the two conflicting measurements of $a_{nn}$: $\ann=-18.7\pm0.7~$fm~\cite{GO99,GO06} and $\ann=-16.1\pm0.4~$fm~\cite{HU001,HU002} occupies much of the rest of this section. If both data points are consistent with the LO prediction (\ref{eq.res.ann}) within the uncertainty of that calculation, a higher-order analysis is needed before any conclusion can be drawn. That theoretical uncertainty results from two expansions and their respective truncation. The error from
expanding the wave function in a finite-dimensional variational space (see above) is discussed in Appendix~\ref{sec.rrgm.numstab} and is found to be $\pm 1.5~$keV there. In Subsections~\ref{subsec.un1} and \ref{subsec.un2} 
the effect of omitted higher-order interactions in the expansion of the Lagrangean shall be analyzed. The additional
contributions result from the broken isospin symmetry of the underlying theory in both the strong and the electromagnetic
sector. The omitted effects in the EFT are divided here into short-range parts, which all occur in the strong part of the charge-symmetry breaking Hamiltonian, and long-range effects, the most important of which are electromagnetic interactions beyond Coulomb repulsion and the impact of the neutron-proton mass difference on $\Delta(3)$. 

%----------------------------------------------------------------------------------------------------------------------
%----------------------------------------------------------------------------------------------------------------------

\subsection{Theoretical uncertainty I (short-range)}\label{subsec.un1}
\par
In this subsection the effect of omitted higher-order NN operators in the Lagrangean which correspond to strong interactions will be assessed. These operators must violate charge symmetry, or they will not contribute to $\Delta(3)$. They become relevant for modes with momenta of order $m_\pi$. In this bound-state calculation these high-energy modes can only affect observables through loops. Their effects are assessed
here in two ways. First, by a change in the regulator parameter $\Lambda$; specifically, $\Lambda=400,800,1600~$MeV and thereby
three different short-range potentials are used. Second, by a change from $\anpt$ to $B(d)$ as input for $C_2$, which
demonstrates how imposing renormalization conditions at different characteristic
momenta changes the output predictions. 
\par
 The LEC values corresponding to these four different cases ($\Lambda=400,800,1600~$MeV, $\anpt$ instead of $B(d)$ as input)
are different because each set represents a different model for the short-distance part of the interaction. The substitution $\anpt \rightarrow B(d)$ was made for $\Lambda=700$~MeV only, and did not result in a significant shift of the
correlation line.
All four sets of LECs reproduce, by construction,
the same long-distance behavior, as given by the specific input data, within higher-order uncertainty. The difference in predictions derived from the four potentials is thus the uncertainty due to permitted variations in the short-distance physics. This sets a lower bound for the
theoretical uncertainty. 
\par
Our assumption that three-nucleon forces do not contribute to $B(t)-B(^3\text{He})$ at leading order is supported by the absence of any significant cutoff variation in our result for this observable.
This cutoff dependence is quantified by the width of the correlation band shown in
Fig.~\ref{fig.res.db3}.
Naive dimensional analysis suggests that the three-nucleon contribution to this binding-energy difference should be less than that of CSB two-nucleon operators,
and the results found here do nothing to contradict that view.
Similar conclusions have recently been reached by other authors~\cite{HK11}. Of course, naive dimensional analysis is strongly violated in \eftnopi~in the isospin-symmetric sector, with the TNI being of leading order there. Thus a power counting for isospin-violating operators in \eftnopi, {\it e.g.} along the lines laid out for sub-leading isospin-symmetric forces in Ref.~\cite{BB04}, remains an interesting open problem. 
\par
\begin{figure}
\includegraphics[width=0.7\columnwidth]{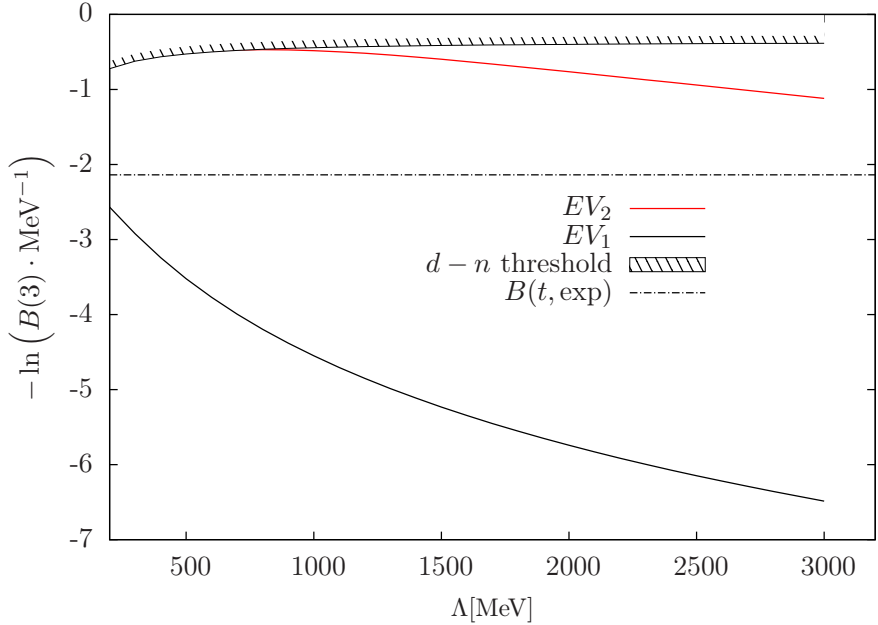}
\caption{\label{fig.lcyc}\small The two lowest eigenvalues of the $nnp$ system of a Hamiltonian with the two-body
interaction fitted to $\anps$ and $\anpt$ without a three-body term, as a function of the regulator width $\Lambda$.}
\end{figure}

The following brief discussion explains the choice of $\Lambda=1600$ MeV as an upper bound for the cutoff.
$\Lambda$ is not increased further because of the corresponding increase in the unrenormalized three-nucleon
binding energy. A diverging three-body ground-state energy is the result of a two-body interaction whose range
is decreased and strength increased to fit $\anps$ and $\anpt$ (Thomas effect). The discrepancy in
the binding energy with and without a three-nucleon force---introduced to properly renormalize one three-nucleon bound-state energy to that 
of the triton---increases in consequence, with more bound states entering the spectrum at specific cutoff values.
In Fig.~\ref{fig.lcyc} the relevant part of the three-nucleon spectrum that is obtained without a TNI is shown as a function of the
regulator. The NN interaction fits $C_{1,2}$ to the singlet and triplet $np$ scattering lengths for each $\Lambda$,
and a new ``Efimov'' trimer is found to enter around $\Lambda=750~$MeV. 
The three-body force has then either to be strong enough to lift a very deeply bound three-body state (solid black line in Fig.~\ref{fig.lcyc}) to the experimental
triton energy while unbinding the other states, or it must pull the shallowest of those states (solid red line in Fig.~\ref{fig.lcyc}) to the triton level. In either case the variational basis must be refined, either to expand states of considerably larger
binding energy, or to treat excited states accurately. Convergence for either case requires a larger RGM space
compared to the one where only one three-body state is bound, and bound with an energy already of the same order of magnitude as that of the
triton. 

For the three representative cutoff values
used to obtain Fig.~\ref{fig.res.db3}, the three-body parameter was always adjusted to elevate the ground state to the triton level, because
this repulsive force pushes all the shallow states that are not typically resolved by the RGM basis above
the $d-n$ threshold.(In the qualitative analysis of the previous paragraph, whose results were shown in Fig.\ref{fig.lcyc}, a small space was chosen
that expands only one of the shallow states, EV$_2$ in Fig.~\ref{fig.lcyc}.) 
The long-distance properties of the three-body state are correctly reproduced by this basis, as witnessed by the fact that
the $d-n$ threshold approaches its predicted value of $B(d)=1/(\anpt^2\mn)\approx1.41~$MeV for $\Lambda\to\infty$.
(The value here is the correct one, given that $\anpt$ is taken as input. Slightly different thresholds will be found at any finite $\Lambda$, since in that case higher-order terms in the effective-range expansion are not zero.)

%---
%---

Bearing this limitation in mind, the graph in Fig.~\ref{fig.res.db3} allows the theoretical uncertainty at this order to be estimated as the maximal $\ann$ difference at the experimental binding-energy splitting:
\begin{equation}\label{eq.res.unc}
\delta_\text{\tiny short}=\frac{1}{2}\vert\ann(800~\text{MeV})-\ann(1.6~\text{GeV})\vert\approx 2~\text{fm}.
\end{equation}
Accordingly, we identify the mean value of all regulator-dependent $\ann$ values with the \eftnopi~prediction. We stress that this choice is somewhat arbitrary
and does not coincide with the choice made in~\cite{Ki10} for the average charge radius of the triton. For that radius, the \eftnopi~prediction was taken to be an
extremal value calculated for the smallest employed cutoff. That choice was made because another LO calculation predicting the triton charge radius~\cite{PL07} showed that the variation of
 cutoff and input data in~\cite{Ki10} was not reflecting the total LO uncertainty.  In fact, both here and in~\cite{Ki10} we take the mean
of all available \eftnopi~predictions for the observable of interest (here, $\ann$~and in~\cite{Ki10} the triton charge radius), regardless of the numerical method used to obtain them. The LO uncertainty is then given as half the difference between the minimum and
maximum value predicted for that observable. This seems a sensible general prescription for defining central values, and uncertainties due to short-distance physics, in \eftnopi~calculations.
\par
Here, only the $\ann=-18.7~$fm data point lies within this uncertainty range.
This result, however, does not yet allow for the conclusion that the smaller datum is inconsistent
with the input data, $\{\anps,B(d),B(^3\text{He}),\app\}$. Only if the uncertainty due to long-range contributions does not add to $\delta_\text{\tiny short}$ an
amount that would increase the total uncertainty to eventually include the second data point, can such a discrimination be made.

\subsection{Theoretical uncertainty II (long-range)}\label{subsec.un2}
As a cutoff variation assesses only the dependence on short-distance structure, an estimate of the theoretical uncertainty is incomplete without
considering effects sensitive to low-momentum modes. Above it was argued why the long-distance part of the interaction can be approximated by
the Coulomb potential.
It is shown here that uncertainties from thereby omitted higher-order interactions, which are {\it not} assessed by the $\Lambda$ variation considered in the previous subsection, are sufficiently
small to render only one experimental data point for $\ann$ consistent with the input observables.
\par
For this purpose, $\Delta(a)$ is treated as a function of $\Delta(3)$, \textit{i.e.}, an interpolation of the dependence whose graph is shown in Fig.~\ref{fig.res.db3} is inverted. Furthermore,
the assumption that higher-order interactions will yield a correlation line which might be shifted but is identical in shape compared to the ones shown is made.
Then, the error in $\Delta(a)$ introduced by the suppressed terms who contribute to $\Delta(3)$ a correction $\Delta_c(3)$ can be approximated via
\begin{equation}\label{eq.error-lr}
\delta_\text{\tiny long}\approx\frac{\partial\Delta(a)\left(\Delta(3)\right)}{\partial\left(\Delta(3)\right)}\Bigg\vert_{\Delta(3)=0.756~\text{MeV}}\hspace{-1.4cm}\cdot\Delta_c(3)\;\;\;.
\end{equation}

The two long-distance effects we must consider are:
\begin{enumerate}
\item The difference in the nucleon kinetic energy due to non-equal neutron and proton masses. Parametrically we estimate the size of this effect to be:
\begin{equation}
\langle \frac{p^2}{2 m_N} - \frac{p^2}{2 m_n} \rangle \approx (m_n-m_p) \langle \frac{p^2}{2 m_N^2} \rangle.
\end{equation}
Taking $\langle \frac{p^2}{2 m_p} \rangle \sim B(t)$ we find:
\begin{equation}
\Delta_c(3;m_n-m_p) \sim (m_n-m_p) \frac{B(t)}{m_N} \sim 10~{\rm keV}.
\end{equation}
This is in good agreement with the value from from Ref.~\cite{KO05}, $\Delta_c(3;m_n-m_p)=14$~keV.

\item The electromagnetic interaction between nucleon magnetic moments, between the currents associated with moving protons, and due to vacuum polarization.
Those effects were calculated numerically to \textit{increase} the mass difference by less than $30~$keV (see \textit{e.g.}~\cite{Pi01} where the various contributions
where calculated with the AV18/IL2 model).
Furthermore, the corrections to the Coulomb potential due to the proton's finite size have to be accounted for.
In fact, these turn out to be the largest of the higher-order electromagnetic effects, since they modify
the Coulomb energy by a fractional amount $\sim r_p^2/R^2$, where $R$ is the typical distance scale which dominates the Coulomb energy. Putting in $r_p \approx 0.85$ fm, $R \approx 2.5$ fm we might expect up to a 10\% effect. Numerical evaluations~\cite{WIS91} however suggest a somewhat smaller number, \textit{reducing} the mass difference by about $33~$keV.
Importantly, the finite proton size decreases the impact of the electromagnetic interaction on the binding energy (thus increasing B(${}^3$He)) while the other effects listed above increase it (and so decrease  B(${}^3$He)). Thus,  each class of correction could individually induce an error of order $30~$keV in our LO binding energy calculation, but they work in opposite directions, such that we can confidently say that their combined effect will not produce more than a $28~$keV shift in $\Delta(3)$.
\end{enumerate}

Combining these two higher-order effects linearly with the
$2~$keV RGM uncertainty (see Appendix~\ref{sec.rrgm.numstab}) we find a potential higher-order correction due to long-distance effects which could be as large as $\Delta_c(3)=44~$keV. Employing Eq.~(\ref{eq.error-lr}), this produces an uncertainty of
\begin{equation}\label{eq.error-lr-value}
\delta_\text{\tiny long}=2.1~\text{fm}.
\end{equation}
The combined theoretical uncertainty in $\Delta(a)$ due to higher-order long- and short-range interactions is then, conservatively, taken to be
\begin{equation}\label{eq.error-tot}
\delta_\text{LO}=\delta_\text{\tiny long}+\delta_\text{\tiny short}=4.1~\text{fm}.
\end{equation}
The central value for $\ann$~between the maximum adopted at $\Lambda\to\infty$ (see eq.~(\ref{eq.res.ann})) and the minimal prediction at $\Lambda=800~$MeV is $-22.9~$fm.
\eftnopi~thus yields, at leading order, a neutron-neutron scattering length of
\begin{equation}\label{eq.ann-pred}
\ann\left(\text{\eftnopi}\right)=-22.9\pm4.1~\text{fm}.
\end{equation}
Hence, the datum $\ann=-18.7\pm 0.7~$fm is consistent with the input data set $\{\anps,\app,B(d,t,^3\text{He})\}$ while the other datum $\ann=-16.1 \pm 0.4$ fm is inconsistent.

\subsection{The limit $\ann\to\app$}\label{subsec.hypo}
\begin{figure}
\begin{minipage}[b]{.49\textwidth}
\centering
\includegraphics[width=0.97\columnwidth]{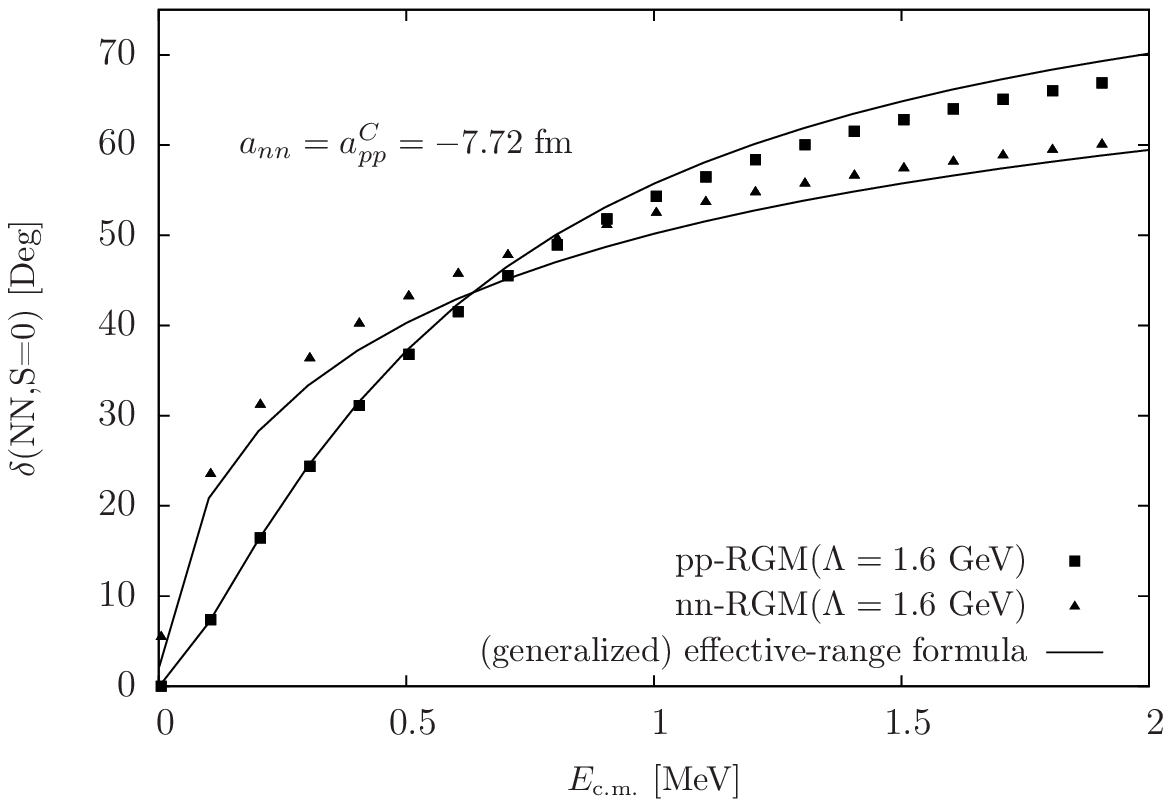}
\caption{\label{fig.res.dnnpp-e}{\small Comparison between the analytic (generalized) effective range formula (\cite{gen-ere}) and the
cut-off ($\Lambda=1.6~$GeV) RGM predictions with \eftnopi~for the ($pp$) $nn$ $^1S_0$ scattering phase shift with $\ann=\app$.}}
\par\vspace{0pt}
\end{minipage}
\hfill
\begin{minipage}[b]{.49\textwidth}
\centering
\includegraphics[width=0.97\columnwidth]{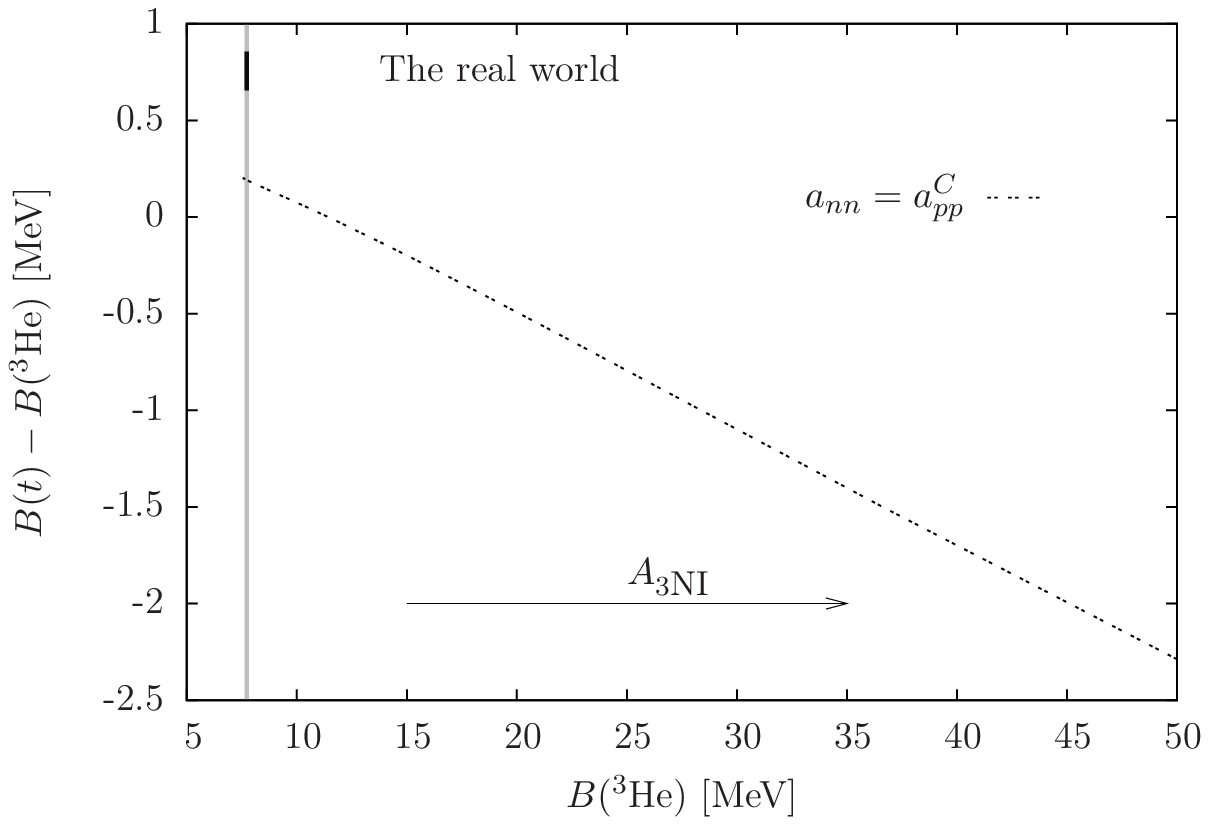}
\caption{\label{fig.res.db3-bhe}{\small Difference between binding energies for the mirror nuclei $t$ and $^3$He as a
function of the $^3$He binding energy at $\ann=\app$.}}
\par\vspace{0pt}
\end{minipage}
\end{figure}

The interactions which generate $\ann$ and $\app$ are different: the short-range part is of the same structure but different in strength. In the $nn$ case
it is solely responsible for $\ann$, whereas, for $pp$, it complements the Coulomb force to yield $\app$. It is therefore not obvious that for
$C^{nn}$ such that $\ann\approx\app$, the triton will be bound by the same amount as $^3\text{He}$ is.
Calculation at $\Delta(a)=0$ (i.e. $\ann=\app$) results in
\begin{equation}\label{eq.res.bt}
\Delta(3)=-0.11\pm 0.1~\text{MeV},
\end{equation}
\textit{i.e.}, two three-body systems of equal binding energy. The bound states result from the same short-range $np$ interaction,
but a purely short-range force between the like pair in the triton, and a combination of a similar short-range counter term plus
the Coulomb interaction in Helium-3. The resultant approximate degeneracy in the binding energies is a reflection
of the fairly small characteristic momenta.
Comparing the $nn$ and $pp$ phase shifts resulting from interactions with $\Delta(a)\approx 0$ at $\Lambda=1.6~$GeV
(Fig.~\ref{fig.res.dnnpp-e}), the $nn$ interaction is found less repulsive than the $pp$ one below $\ecm\approx 0.65~$MeV, equal around $0.65~$MeV,
and more repulsive for $\ecm\gtrsim 0.65~$MeV. If bound states receive significant contributions from modes with $k_\text{\tiny c.m.}\lesssim\sqrt{\mn (2~\text{MeV})}=:p_\text{\tiny balance}\approx45~$MeV,
then the difference in $V_{nn}$ and $V_{pp}$ is na\"ively expected to balance, yielding approximately the same binding energies, as we see in Eq.~(\ref{eq.res.bt}), since $\ann\approx\app$ produces $\Delta(3)\approx 0$. 
In fact, this analysis implies a momentum distribution amongst the nucleons within the triton bound state which is dominated by momenta markedly smaller than the conventional estimate: $p_\text{\tiny typ}:=\sqrt{2\cdot 2/3\mn\cdot B(t)}\approx100~$MeV---at least as far as the momenta pertinent to the binding-energy difference $\Delta(3)$ are concerned. 
\par
Furthermore, the reasoning implies that more deeply bound mirror nuclei will exhibit a larger difference in their binding energies, even though the
respective two-body scattering lengths are equal. In a world where not only $\Delta(a)=0$ but additionally $B(^3\text{He})\gg B(^3\text{He,exp})$, the uncharged mirror image would be
not as deeply bound, $B(t)\ll B(^3\text{He})$, as a result of the stronger repulsion of the uncharged `neutrons' as compared to protons at relative momenta greater than about 20 MeV. 

This hypothesis is confirmed by the results of a RGM calculation (Fig.~\ref{fig.res.db3-bhe}). A smooth change of the three-nucleon parameter $\atni$
increases $B(3)$ but leaves the two-nucleon sector invariant, \textit{i.e.}, $\ann=\app\approx -7.8~$fm, $\anps\approx -23.75~$fm, and the deuteron at its physical
binding energy. The result is an increasingly less bound uncharged system relative to its charged mirror sibling, as conjectured above.
In more detail, the triton is found not as deeply bound (intersection of dashed line with gray band in Fig.~\ref{fig.res.db3-bhe}) as in our world (black band in Fig.~\ref{fig.res.db3-bhe})
for a TNI producing the physical Helium-3 binding energy. This is a consequence of the more repulsive $nn$ force at low momenta implied by $\Delta(a)=0$. Adjusting the TNI to
yield a more deeply bound Helium-3 widens the gap between its ground state and the $p-d$ breakup threshold because $B(d)$ remains constant. The triton binding energy also
increases in this procedure and is found larger than $B(^3\text{He})$ for $B(^3\text{He}) \approx 12~$MeV ($\Delta(3)>0$ as shown by dashed line in Fig.~\ref{fig.res.db3-bhe})
becoming increasingly less bound relative to $B(^3\text{He})$ as $B(^3\text{He})$ increases further. This behavior of $B(^3\text{He})$ is in accord with the qualitative discussion above. 
\par
%---
%---
For systems with $p_\text{\tiny typ}$ much smaller than $p_\text{\tiny balance}$ the opposite behavior is expected. If we assume, for the moment, that $\ann=\app$, the above reasoning
can be used to shed some light on the situation in the six-body system. In $^6$He, the two halo neutrons are very weakly bound together with an $\alpha$ core. Of course, it is possible that $^6$He would not be bound if $\ann$ were reduced to agree with $\app$. But, even if $^6$He were bound for this smaller $\ann$, the momenta in that bound state are low enough that the mirror nucleus $^6$Be with two `halo' protons will {\it not} be bound. The $pp$ interaction is more repulsive for those low momenta---due to Coulomb effects. An explicit calculation at an order in
\eftnopi~which generates $^6$He as a shallow bound state is required to validate this hypothesis. At present, a leading-order analysis is not able to reach this level of accuracy.

\section{Conclusion}

The Coulomb energy $\langle V_C \rangle$ of the three-nucleon system can be computed reliably with $\eftnopi$ wave functions. In Ref.~\cite{
Ki10} an NLO \eftnopi~computation gave a value of $660 \pm 30~$keV for $\langle V_C \rangle$. In this work we have considered, in addition to Coulomb effects, the impact of the charge-symmetry-breaking NN operators which produce different (strong) $pp$ and $nn$ scattering lengths. We carried out the LO \eftnopi~calculation for three different cutoffs $\Lambda=400$, 800, and 1600 MeV using the modified renormalization-group method. We found a robust correlation between $B(t)-B({}^3\text{He})$ and the difference of scattering lengths $\Delta(a) :=\ann - \app$. The fact that this correlation is largely independent of the short-distance physics in the NN system indicates that three-nucleon operators do not contribute to this isospin-violating difference of binding energies at leading order. 

From the correlation and the experimental values of the tri-nucleon binding energy difference and $\app$ we infer:
\begin{equation}
a_{nn}=-22.9 \pm 4.1~{\rm fm}.
\label{eq.final}
\end{equation}
The uncertainty here has been assessed by adding linearly estimates of the impact of higher-order, short-distance operators in the NN system and of neglected long-range effects ({\it e.g.}, magnetic-moment interactions, as well as the nucleon mass difference). ``Short-" and ``long"-distance  effects of higher order appear to contribute roughly equal amounts to the error bar. 

The result (\ref{eq.final}) is due to operators that are first-order in isospin breaking, and so first-order perturbation theory with these operators, evaluated between charge-symmetric triton wave functions, could also have been employed (c.f.~Ref.~\cite{HK11}). Here we performed an assessment of isospin violation in the Hamiltonian in which the relevant interactions were treated non-perturbatively. Such a calculation is more straightforward technically within the RGM. At the level of accuracy of our calculation, the only operator for which this non-perturbative/perturbative distinction might make a difference would be the $pp$ Coulomb potential. But, even there, we anticipate that the second-order piece of the ${}^3$He Coulomb energy is of the same size as other effects neglected in this calculation.  

In order to refine the constraint (\ref{eq.final}) it will be necessary to compute explicitly higher-order electromagnetic effects and the impact of the proton-neutron mass difference on the binding energies. Next-to-leading order and next-to-next-to-leading order triton wave functions in the charge-symmetric sector~\cite{Bd02,Ki10} should be considered. Analysis along the lines of Ref.~\cite{BB02} would also be needed so as to determine the order at which charge-symmetry-breaking three-nucleon operators enter the \eftnopi~calculation. 

\section*{Acknowledgements}
This work was supported by the US Department of Energy (Office of Nuclear Physics, under contract DE-FG02-93ER40756 with Ohio University). 

\appendix
\section{Numerical stability}\label{sec.rrgm.numstab}
The RGM is used to fit LECs and for predictions in the two- and three-body sector. In this section two analyses are presented
to estimate the numerical uncertainty: the convergence of a $B(t)$ calculation with respect to dimension and ``quality'' of the
variational basis for a given set of LECs, and the dependence of $C^{pp}_S$ on the parameters used to expand and regulate
Coulomb functions in a Gaussian basis.
\par
\begin{figure}
\includegraphics[width=0.9\columnwidth]{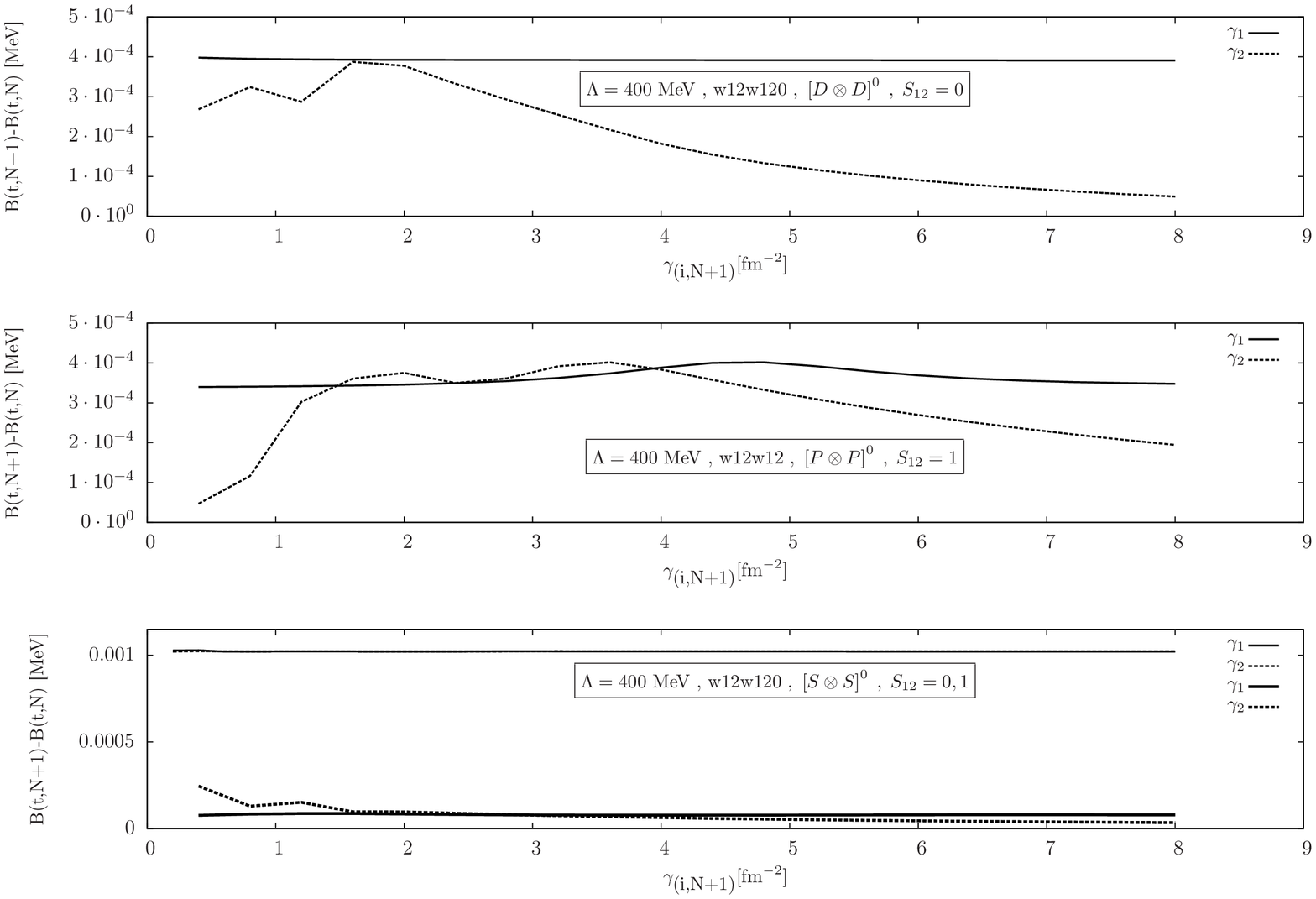}
\caption{\label{fig.num-bt}\small Difference in $B(t)$ due to the addition of basis vectors to a reference set - w12w12(0).
This change is shown in the lowest panel as a function of either of the two width parameters $\gamma_{1,2}$ for a d-n
(\mbox{d\hspace{-.55em}$^-$}-n) grouping in two S waves. The middle (top) panel displays the change for the nucleons in relative P (D) waves.}
\end{figure}

In the first scenario, the interaction is specified through five LECs in Eq.~(\ref{eq.pot-vertex}):
$C_{1,2},C^{nn,pp}_S,C_{3\text{NI}}$.
Predictions for $B(t)$ will depend on the dimension of the RGM basis, $D=D_1+D_2+D_3$, and a ``wise'' choice of width parameters,
$\left\lbrace\gamma_{ij},j=1\ldots D_i\right\rbrace$, for each grouping $i$. We consider ourself wise because the widths are chosen to
expand an object of limited size, which is estimated by the deuteron and triton binding energies to be described within a central potential
whose range is set by the regulator cutoff $\Lambda$. A larger $\Lambda$ relates to a shorter-range interaction, mandating larger
widths $\gamma_{ij}$ to account for the larger values of the wave function resulting from the deeper well. Simultaneously, the exponential tail
has to be modelled accurately by keeping the smaller widths corresponding to the longer-range part. In essence, larger $\Lambda$s require
larger bases but do not pose an in-principle limitation for the application of the RGM. To assess whether a certain variational basis expands
the triton accurately, the supposedly complete basis, $\left\lbrace\vert i\rangle,i=1,\ldots,D_i\right\rbrace$ with $D_{1,2}=60$ and $D_3=0$,
is extended by 5 vectors
all taken from the dominant grouping---which, for the triton, is the deuteron-neutron one. In the new, ($D'=(D_1+5)+D_2+D_3$)-dimensional space, the triton
binding energy is calculated as a function of one width parameter, either $\gamma_1$ for the deuteron fragment, or $\gamma_2$ for the
separation of the neutron from the deuteron. Figure~\ref{fig.num-bt} displays the graph of the function
$f\left(\gamma_i\right)=B(t,D')-B(t,D)$, with $B(t,X)$ being the smallest eigenvalue of the system
\begin{equation}\label{eq.gen-ev-prob}
\big\langle\phi_m\big|\hat{H}\big|\phi_n\big\rangle
=E\big\langle\phi_m\big|\phi_n\big\rangle\;\;\;,
\end{equation}
with indices $m,n$ specifying the $X$ variational parameters $c_{ij}$ in Eq.~(\ref{eq.gauss-exp-coord-wfkt}).
%----------------------------------------------------------------------------------------------------------------------
%----------------------------------------------------------------------------------------------------------------------
The numerical uncertainties due to the finite basis are summarized in table~\ref{tab.num-unc}.
%--
\begin{table*}
\renewcommand{\arraystretch}{1}
  \caption{\label{tab.num-unc}{\small Uncertainty in the triton binding energy $\Delta B(t)$ due to the
omission of specific basis states, \textit{e.g.}, an addition of vectors of the deuteron-neutron grouping with particles in relative S waves
and support from approximately $1/\sqrt{8}~$fm to infinity will change $B(t)$ by less than $0.0011~$MeV.}}
\footnotesize
\begin{tabular}{c|c|c|c}
\hline
cluster $\left[s_n\otimes s_p\right]^{S_{12}}$ & orbital angular momentum $\left[l_1\otimes l_2\right]^L$ & Gaussian & $\Delta B(t)~$[MeV]\\
\hline\hline
d-n $S_{12}=1$ & $\left[0\otimes 0\right]^0$ & $\gamma_{1,2}\in[0,8]~\text{fm}^{-2}$ & 0.0011\\
\hline
\mbox{d\hspace{-.55em}$^-$}-n $S_{12}=0$ & $\left[0\otimes 0\right]^0$ & $\gamma_{1,2}\in[0,8]~\text{fm}^{-2}$ & 0.0003\\
               & $\left[1\otimes 1\right]^0$ &                                       &  0.0003\\
               & $\left[2\otimes 2\right]^0$ &                                       &  0.0003\\
\hline
    \end{tabular}
\end{table*}
%--
In conclusion, a total uncertainty in
the three-body binding energy $B(3)$---the numbers are of the same order of magnitude for ${}^3$He---due
to the truncation of the variational basis of
\begin{equation}\label{eq.num}
\Delta\left(\text{RGM}\right)=\pm 1.5~\text{keV}
\end{equation}
is assigned to this analysis. This value is markedly less than the na\"ive $10~$\% LO \eftnopi~of $B(t,\text{exp})$, and hence the
120-dimensional basis is sufficient for the accuracy of this order.
\par
The LEC values depend on the loop regulator $\Lambda$ and on the RGM basis that spans the space in which they are fitted to data.
For a cutoff variation at $\Lambda>m_\pi$ and a modification of the basis using states with support only for particle separations
less than approximately $1/m_\pi$, both dependencies reflect a modification of non-observable high-energy modes absorbed in the LECs.
However, the dependency on the model space is sought to be minimal, so as to allow application of and comparison with the LEC values found using other
numerical methods. To assess this uncertainty, the 20-dimensional S-wave two-body basis used throughout this work was refined in two ways:
\begin{itemize}
 \item add one basis state and determine $C^{pp}_S$ as a function of the width parameter;
 \item determine $C^{pp}_S$ as a function of the regulator parameter used for the irregular Coulomb function;
\end{itemize}
In both cases, the change in the LEC was $-0.01<\frac{\Delta C^{pp}_S}{C^{pp}_S}<0.001$, and the resulting
effect on two- and three-body observables relevant for this work was small relative to the anticipated leading-order \eftnopi~accuracy.

\end{document}